\documentclass[12pt]{iopart}

\usepackage{iopams}  
\usepackage{graphicx}

\begin{document}

\title[Coarse-grained modeling of protein unspecifically bound to DNA]
{Coarse-grained modeling of protein unspecifically bound to DNA}

\author{
Carlo Guardiani\,$^{1}$
Massimo Cencini\,$^2$
and Fabio Cecconi\,$^2$
%\footnote{To whom correspondence should be addressed.
%Tel: +39 06 49937453; Fax: +39 06 49937440;}
}
\address{$^{1}$Dipartimento di Fisica,
  University ``Sapienza'', Piazzale A. Moro 2, 00185 Rome Italy\\
$^{2}$Istituto dei Sistemi Complessi,
  Consiglio Nazionale delle Ricerche, via dei Taurini 19, 00185 Rome,
  Italy}

\ead{fabio.cecconi@roma1.infn.it}

\begin{abstract}
There is now a certain consensus that Transcription Factors (TFs)
reach their target sites, where they regulate gene transcription, via
a mechanism dubbed \emph{facilitated diffusion} (FD). In FD, the TF
cycles between events of 3D-diffusion in solution (jumps),
1D-diffusion along DNA (sliding), and small jumps (hopping) achieving
association rates higher than for 3D-diffusion only.  We investigate
the FD phenomenology through Molecular Dynamics simulations in the
framework of coarse-grained modeling. Despite the crude
approximations, we show that the model generates, upon varying
equilibrium distance of the DNA-TF interaction, a phenomenology
matching a number of experimental and numerical results obtained with
more refined models.  In particular, by focusing on the kinematics of
the process, we characterize the geometrical properties of TF
trajectories during sliding.  We find that sliding occurs via helical
paths around the DNA helix leading to a coupling of translation along
the DNA-axis with rotation around it. The 1D-diffusion constant
measured in simulations is found to be interwoven with the geometrical
properties of sliding and we develop a simple argument able to
quantitatively reproduce the measured values.
\end{abstract}

\vspace{2pc}
\submitto{\PB}
\maketitle

\section{Introduction}

Transcription Factors (TFs) play a key role in the regulation of gene
expression acting as gene-transcription activators or inhibitors both
in prokaryotes and eukaryotes~\cite{Alberts}.  One of the most
fundamental issues in protein-DNA recognition is the ability of TFs to
selectively identify their specific target sites that are embedded
among tens of millions of competing non-specific DNA sequences. A
related issue pertains to the high rate of recognition of the specific
target sites. As early as 1970, Riggs et al.~\cite{Riggs70} observed that the
\emph{lac} repressor in \emph{E. coli} can associate to the cognate
operator sequence at a rate about two orders of magnitude higher than
that predicted by Smoluchowski equation for a diffusion-limited
association reaction.

Berg, Winter and von Hippel~\cite{Berg81} explained this paradox
suggesting that TFs do not target their sequences through pure
3D-diffusion but they can also diffuse while being unspecifically
associated (mainly due to electrostatic interactions \cite{halford2004})
to the DNA. Such a dimensional reduction, 
dubbed \emph{facilitated diffusion} (FD), can make the search more 
efficient speeding up the identification of target sites.  
More specifically, FD proceeds by
means of four pathways \cite{Berg81}: $(i)$ \emph{sliding} along the
DNA, $(ii)$ \emph{hopping}, $(iii)$ \emph{jumping} and $(iv)$
\emph{intersegmental transfer}.  During sliding, the TF remains
in unspecific contact with the DNA chain performing monodimensional
diffusion along its contour. During hopping, the TF detaches from the
DNA but reassociates with it at a short distance from the dissociation
point.  During jumping, the TF dissociates from the DNA undergoing
free 3D-diffusion and rebounds to the DNA in a completely uncorrelated
location. Finally, in the intersegmental transfer, relevant to compact DNA
conformation, the TF transiently binds two non-contiguous DNA branches
allowing its transfer from one DNA segment to the other. The latter process 
requires the possibility for the TF to bind at multiple loci. 

Facilitated diffusion has been extensively studied through analytical
models~\cite{Berg81,halford2004,Slutsky2004,Sheinman2012} which
achieve closed-form solutions at the price of a drastic simplification
in the complexity and the heterogeneity of the genome.  The
approximation of the TF-DNA affinity landscape, for instance, may lead
to significant deviations from the experimental patterns.  A more
detailed level of description is based on computational stochastic
models which allow large-scale simulations involving DNA stretches of
the order of $10^{6}$bp, tens of thousands of TFs and can reach the
time-scale of a few seconds~\cite{Chu2009,Zabet2012}.  This high
performance, however, relies on a set of assumptions that are considered
quite controversial and that need further elucidation.  More
specifically, the issues include (i) the proportion of sliding and
hopping during 1D-diffusion; (ii) the fraction of time the TF spends
in 3D- and 1D-diffusion; (iii) the effects of molecular crowding
related to the presence of multiple copies of the TF that prevent each
other's movement acting as moving roadblocks.

In order to clarify these issues, experimental studies can be
profitably integrated with coarse-grained molecular simulations. For
instance, while fluorescence experiments have allowed the direct
observation of a single TF moving along DNA (confirming the
facilitated diffusion theory)~\cite{Elf2007}, the spatial resolution
of the technique does not discriminate hopping from sliding.  Thus a
quantitative characterization of the two kinds of motion still remains
elusive. Another source of ambiguity concerns the values of the mono-
and tri-dimensional diffusion constants.  While there is a general
consensus on the fact that $D_{1} < D_{3}$, the measured values of
these constants vary by several orders of magnitude according to the
particular DNA sequence and the experimental
set-up~\cite{Wang2006,Graneli2006}.  This variability is anything but
irrelevant since it is closely related to the so-called
\emph{speed-stability paradox}~\cite{Slutsky2004,Benichou2008}.  In
fact, on the one hand, a high diffusion constant allows a fast
scanning of non-specific sites improving the search of the target
sequences, on the other hand, high $D_{1}$ values can only be attained
at the price of a low TF-DNA affinity that may destabilize the complex
formed by the TF with its specific target site.

Another problem that has not yet come to a conclusive answer is the
fraction of time spent by the TF in 3D-diffusion and in sliding.
Assuming that only sliding and jumping are at work, simple analytical
arguments \cite{halford2004,Slutsky2004} suggest that the average time
necessary to reach the target is $ t_{s} =
(\tau_{1}+\tau_{3})M/\overline{n}$, where $M$ is the total number of
sites, $\overline{n}$ is the average number of sites scanned during a
sliding event and $\tau_{1}$ and $\tau_{3}$ are the average durations
of individual episodes of sliding and 3D-diffusion, respectively.
Assuming that the search time has been to some extent optimized 
  by evolution, $t_{s}$ is minimal if $\tau_{1} = \tau_{3}$, \emph{i.e.} 
  when the TF spends exactly the same amount of time in sliding and
  3D-diffusion. This hypothesis of optimality, however, contrasts with
  experimental studies in bacteria suggesting that the TF spends much
  more time in sliding than in 3D-diffusion
  ($\tau_{1}/(\tau_{1}+\tau_{3}) = 0.9$)~\cite{Elf2007}. Even though
  the discrepancy may be due to the absence of hopping in the above
  argument, of course, one cannot exclude that evolution has selected a
  suboptimal solution.

All of these problems can be addressed through Molecular Dynamics
simulations but unfortunately not in the framework of atomistic
methods. The longest atomistic simulation on DNA reported to date has
been a few microseconds \cite{Laughton2011} while sliding events
typically involve timescales of $O(s)$ and sliding length of
$O(100)$bp~\cite{Mirny2009}. Thus, it is clear why atomistic MD is
unfit to the study of facilitated diffusion and the resorting to a
coarse-grained phenomenological modeling is mandatory.  Recently,
Brackley et al. \cite{Brackley2012} introduced a coarse-grained model
portraying the TF as a sphere with a binding site on its surface and
the DNA as a chain of beads. The model, also accounting for both DNA
flexibility and sequence heterogeneity, showed that the search time
could be minimized by an appropriate tuning of the TF-DNA
affinity. Givaty and Levy~\cite{Givaty2009} proposed a much more
detailed model whereby DNA is simplified as a double stranded helix
with three beads per nucleotide, while the TF is described as a bead
for each residue. Levy's simulations show that during sliding the TF
remains deeply buried into the major groove and presumably makes use
of the same binding interfaces for both specific and non-specific DNA
interactions.

\begin{figure}[t!]
\centering
\includegraphics[width=0.5\textwidth]{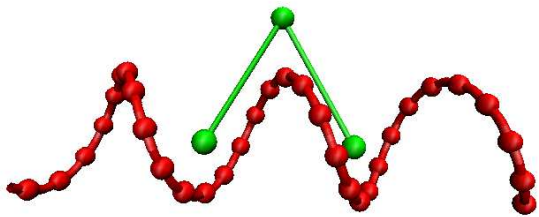}
\caption{Cartoon representation of the TF (in green) and DNA (in red)
  model used in this work, see text for details. \label{fig:0}}
\end{figure}

In this work, we introduce and study a model with a level of resolution
intermediate between those mentioned above. With reference to 
Fig.~\ref{fig:0}, a TF is portrayed as a triangular object with the first and
last bead representing the DNA binding regions, so to mimic the basic
features of homodimeric prokaryotic TFs. The central bead models
the scaffold of the protein imparting the correct orientation to the
DNA-binding domains.  The DNA is represented by a single helix frozen in 
its equilibrium conformation as it greatly facilitates the identification 
of the various searching regimes.
%----------------------------------------------------------------
%----------------------------------------------------------------

The aim of our work is to develop a toy model including as few
ingredients as possible, yet able to capture the known phenomenology of
TF dynamics unspecifically bound to DNA.  The approach enabled us to
connect the mono-dimensional diffusion coefficient to the geometrical
properties of the TF trajectories confirming previous results
\cite{Bagchi2008,Schurr1979}.  This suggests that two minimal
key-elements are sufficient: the helical topology of DNA and a
confining DNA-TF interaction tethering the TF in the neighborhood of
DNA.

We restricted our study to the case of purely non-specific TF-DNA
interaction whose importance cannot be underestimated, as it is
reasonable to assume that a TF before reaching its target spends most
of the time in non-specific attraction with DNA sites. On the other
hand, this is at the core of the searching process, which being
``unproductive'', requires speeding mechanisms or optimal pathways to
be shortened.

A preliminary exploration of parameter space has been performed to
obtain behaviors that reasonably matched the principal features of
the facilitated diffusion phenomenology. Then we run simulations to
analyze the TF dynamics in the DNA proximity in order to characterize
the interplay between three searching modes: sliding, hopping, and
free diffusion.  As we shall see, our simulations show the existence
of these three different regimes whose mutual prominence depends on
$\sigma$, the minimum of the Lennard-Jones (LJ) potential, that
dramatically affects the energy landscape.  Specifically, we found
that sliding occurs via the coupling of rotation and translation along
the DNA, in which the TF propagates one- dimensionally along the DNA
while rotating along the DNA-helical contour.  This characteristic
motion is consistent with experimental observations for several
proteins~\cite{Gorman_REV}.

Finally we have also quantified how each mechanism contributes to
search efficiency.

%%%%%%%%%%%%%%%%%%%%%%%%%%%%%%%%%%%%%%%%%%%%%%%%%%%%%%%%%%%%%%%%%%
\section{MATERIALS AND METHODS\label{Sec:methods}}
%%%%%%%%%%%%%%%%%%%%%%%%%%%%%%%%%%%%%%%%%%%%%%%%%%%%%%%%%%%%%%%%%%

\subsection{Model\label{sec:model}}

In the following we briefly describe the coarse-grained representation
used in this study for the DNA chain and the transcription Factor.

\emph{DNA helix\quad} During the simulations the DNA was kept frozen
in its initial configuration so that DNA beads, used to represent DNA
bases, do not interact with one another.  The DNA configuration is
chosen to be a straight helix as, unlike Ref.~\cite{Brackley2012}, we
want to retain the helical geometry in order to understand its effect
on the FD process.  However, since we adopt a very crude model for the
TF (see below), we do not need the detailed description of the DNA
double helix proposed in Ref.~\cite{Givaty2009}. We thus consider a
minimal representation in terms of a single straight helix where each
bead represents a base-pair (bp). In particular, to mimic the typical
conformation of B-DNA \cite{Alberts} we consider an helix of radius
$\varrho = 13.0$~\AA\; with $h=10.5$ base-pairs per helix turn and the
distance between two consecutive bases along the DNA axis (here chosen
to be along $z$) is taken to be $b=3.32$~\AA\,, so that the helix
pitch is $P=hb=34.86$~\AA\,.  In this way the coordinates of the
$n$-th bead are simply obtained from the parametric equations of the
helix
\begin{equation}
 \begin{array}{ll}
x(n)  = & \varrho\cos(2\pi z(n)/P) \\
y(n)  = & \varrho\sin(2\pi z(n)/P), \qquad n=1,\ldots,N\\
z(n)  = & b n\,.
\end{array}
\label{eq:helix}
\end{equation}
The total number of base-pairs $N$ in our simulations is $N=1000$
which is larger than the DNA persistence length (about $100$bp in
physiological conditions \cite{Alberts}). However, assuming a linear
conformation longer than that found in vivo is relevant to single
molecule experiments, where DNA chains are typically stretched (see,
e.g., \cite{Blainey2009}).

\emph{Transcription Factor\quad }The modeling of the TF requires some
discussion. In prokaryotes, TFs are typically homodimeric as they
target palindromic DNA sequences~\cite{BrandenTooze}.  In each
monomeric subunit the DNA recognition region is an helix-loop-helix
motif whereby the second helix is designed to fit into the major
groove establishing hydrogen bonds and hydrophobic interactions with
the nucleotide bases. Since the helix-loop-helix motifs of the two
subunits must fit into two adjacent major grooves, they are located at
the typical distance of one pitch $P\approx 32-34$~\AA.  In
eukaryotes, TFs can be both homodimeric and heterodimeric so as to
increase the range of DNA sequences to be recognized.  For instance,
steroid hormones receptors are typical homodimeric receptors while the
TFs containing the leucine zipper motif are normally heterodimeric and
the helix-loop-helix TF can be both homo- and heterodimeric
\cite{BrandenTooze}. Our modeling approach aims at reproducing the
basic features of prokaryotic TFs. Therefore, we portray the TF as
three beads arranged at the vertices of an equilateral triangle of
side 32~\AA, to roughly fit the distance between two major grooves. A
variation of the side in the range $28-35$\AA\; or isosceles TF
conformations do not affect the essence of the results.  The first and
third bead can be thought of as the centers of mass of the
DNA-recognizing regions of the two subunits.  The third bead
represents the center of mass of the portion of the TF not directly
involved in DNA recognition, which typically stays away from the DNA
helix.

The TF triangular structure is enforced by the following interactions.
The 1-2 and 2-3 distances of the TF beads are allowed to undergo small
oscillations around their equilibrium value, $r_0$, via a stiff
harmonic potential~\cite{Hinchliffe2008}
\begin{equation}
V_{h}(r_{i,i+1}) = \frac{k_{h}}{2}(r_{i,i+1} - r_{0})^{2}\,,
\end{equation}
whereas the distance 1-3 is maintained via a bending potential
\begin{equation}
V_{\theta}(\theta) = \frac{k_{\theta}}{2}(\theta - \theta_{0})^{2} \,.
\end{equation}
Being interested in the phenomenology of FD and not in the target
search time, we assume only unspecific interactions between the TF and
the DNA chain, which are modeled as described below.  Bead 1 and bead
3 interact with the DNA beads through a standard 12-10 Lennard-Jones
(LJ) potential
\begin{equation}
V_{LJ}(r_{ij}) = 5\epsilon \left[ \left( \frac{\sigma}{r_{ij}} \right)^{12} 
- \frac{6}{5} \left( \frac{\sigma}{r_{ij}}\right)^{10}  \right] \,,
\label{eq:lj}
\end{equation}
where $r_{ij}$ is the distance between bead $i \in \{1,3\}$ of the TF
and bead $j$ of the DNA.  The parameter $\epsilon$ determines the
well depth of the LJ potential while $\sigma$ tunes the position of
the minimum. Thus $\sigma$ determines the equilibrium distance of the TF 
from the helix axis, the larger is $\sigma$
the farther is the equilibrium position of TF from DNA. 
Conversely, the bead 2 of the TF interacts with the
nucleotides of DNA through a repulsive, excluded-volume potential
\begin{equation}
V_{rep}(r_{2j}) = \epsilon_2 \left( \frac{\sigma_2}{r_{2j}}
\right)^{12} \,.
\label{eq:vrep}
\end{equation}
This potential forces the central bead of the TF to point away from
the DNA axis imparting the correct orientation to the transcription
factor. 
In our simulation, we keep $\epsilon$ fixed to set the
energy scale and varied $\sigma$ in a wide range of values. 
For the sake of clarity, the list and the 
values of the parameters defining the DNA-TF model are 
summarized in Table~\ref{tab:model}.  
     
{\it Simulation box\quad} Since the focus of our investigation was the
sliding behavior of the TF, we introduced a cylindrical confinement
potential~\cite{Cacciuto2006}:
\begin{equation}
V_{conf} = V_{xy} + V_{z} = \frac{k_{B}T}{(R_{xy}-r)^{2}} +
\frac{k_{B}T}{(R_{z}-|\Delta z|)^{2}}\,,
\end{equation}
In this expression $r=\sqrt{x^2+y^2}$ is the distance between a TF
bead and the DNA axis that was set to coincide with the $z$-axis,
$\Delta z$ is the distance along the $z$-axis between the bead of the
TF and the center of mass of the DNA, $R_{xy}=100$~\AA~ is the radius
of the cylindrical confinement region. The parameter $R_z$ represents
half the height of the confinement region that we set equal to half
the length of DNA plus 1.5 helical turns. The $V_{xy}$ component of
the confinement potential forces the TF to remain in a circular region
of radius $R_{xy}$ centered on the DNA axis while the $V_{z}$
component prevents the TF to exceed a distance equal to $R_{z}$ from
the DNA center of mass along the $z$-axis.  

The value $R_{xy} = 100$\AA~ for the simulation box can be justified
using the following argument. The average volume available to
interphasic-DNA spans the range $10^{11}-10^{12}$ \AA$^3$.  We can
assume that this is the volume of a spherical region $V=4\pi R_g^3/3$
with $R_g$ being the DNA gyration radius. Following Berg and Blomberg
\cite{Berg1977}, we can construct around the DNA contour a coaxial
cylinder with a volume equivalent to the sphere $4\pi R_g^3/3 = \pi
R_{xy}^2 L$, where $L \sim 10^7$ \AA~ is the typical DNA-length. 
This yields values of $R_{xy}$ in the range $60-200$\AA. 
On not too long timescales, the TF may be reasonably assumed to be confined 
in a cylindrical region of radius $R_{xy}$ around a DNA segment.
Since metaphasic DNA is more condensed, it can be assumed to
be confined in a cylindrical region with a smaller radius $R_{xy}$.
In this situation, the TF can be expected to spend a smaller
fraction of time in 3D-diffusion similar to what happens for small
values of $\sigma$ (see Sect.~\ref{sec:Statistics}). 
Moreover the TF will have a greater tendency to rebind the DNA in
the neighborhood of the point of detachment. In this regime there
will be only a weak interplay between sliding and 3D-diffusion,
leading to a low efficiency of exploration of new sites 
(see Sect.~\ref{sec:Efficiency}). 
This appears to be consistent with the fact
that tightly packed DNA is normally not transcribed nor replicated
but rigidly transferred to daughter cell during mitosis.

We performed Langevin Molecular Dynamics simulations using a
stochastic position Verlet integration scheme~\cite{Melchionna2007}
with time step $h = 0.002$ and friction coefficient $\gamma$ (see
Table~\ref{tab:model}). The simulation time unit can be converted
to the physical one by using the time scale $\tau=\sigma_2
\sqrt{m/\epsilon}$ \cite{Padding}. 
With $\epsilon = 4 k_B T \simeq 16 \times 10^{-21}$J, $\sigma_2 = 5$\AA~,
and assuming an average mass $m \simeq 10$KDa for each bead of the TF, 
we obtain $\tau \sim 9$ps. 

As customary, the Lennard
Jones interactions were truncated to a cutoff distance $r_{c} =
4\sigma$ to speed up the calculations.

\begin{table}
\caption{Table summarizing the parameters and their values  
used in the DNA-TF interaction model and in the simulations.\label{tab:model}}
\medskip
\begin{tabular}{|ll|ll|}
\hline
Parameter & Value & Parameter & Value 
\\
\hline
$\epsilon$    & $1$              & $\sigma_2$   & $5$\AA
\\
$\epsilon_2$  & $0.8\epsilon$    & $r_0 $         & $32$\AA 
\\
$k_h$         & $50 \epsilon$    &  $\theta_0 $    & $60^o$
\\ 
$k_{\theta}$    & $20 \epsilon$    & $\gamma$     & $1$  
\\
$k_B T$       & $0.25 \epsilon$  &  $ m $          & $1$  
\\
\hline
\end{tabular}
\end{table}

\subsubsection{Statistical analysis}

\subsubsection{Determination of sliding, hopping and jumping events
\label{sec:methods:statistics}}

As discussed in the introduction, the process of facilitated diffusion
proceeds by means of four pathways \cite{Berg81}: (i) sliding along
the DNA, (ii) hopping, (iii) jumping and (iv) intersegmental transfer.
In our model, due to the chosen conformation for the DNA chain, only
the first three mechanisms are at work. In order to compute the
statistics of sliding, jumping and hopping, it is necessary to define
the criteria discriminating each event, which are described below.

We consider the TF to be in the sliding regime if the closest DNA
neighbor of bead-1 or bead-3 of the TF is below a distance cutoff of
$1.2\sigma$. This criterion allows the identification of a number of
sliding and non-sliding windows. With this criterion it may
happen that the TF is bound to the DNA with only one bead while the
other is detached. We have studied the statistics of such events and
found that when sliding occurs the percentage of time spent in
two-bead sliding is: higher than $90\%$ for $\sigma\geq 8$\AA; between
$40\%$ and $90\%$ for $4$\AA$\leq \sigma<8$\AA; and decreases up to a
few $\%$ for $1\leq \sigma<4$\AA.

For each value of $\sigma$, the average sliding length, $\langle
|\Delta Z_s| \rangle$ is measured as the average distance covered by
TF between an attachment and the first subsequent detachment.
Non-sliding windows are classified as hopping events if the
displacement of the TF along the DNA axis is smaller than twice
$\langle |\Delta Z_s| \rangle$, otherwise the event is considered as
jumping. The idea underlying this choice is that hopping implies
short-range flights between dissociation and reassociation points
\cite{Gowers2005}.

Clearly, the discrimination between sliding and hopping, and hopping
and jumping suffers of a certain degree of arbitrariness due to the
necessity to introduce specific thresholds on distances. However, upon
varying the threshold values, we verified the results are
qualitatively the same but for some quantitative effects on hopping
statistics.

\subsubsection{Computation of the sliding diffusion constant\label{sec:methods:D1}}
The monodimensional diffusion process of the TF during sliding is
characterized by the diffusion coefficient, $D_1$, along the DNA axis.
The constant $D_1$ can be estimated through the trajectories of molecular
dynamics from the mean square deviation (MSD) along the $z$-axis
during the sliding events.  First, the trajectory of the TF is
segmented into sliding, hopping and jumping events as described
above. Second, in each sliding window $w$ one computes the MSD on the
window as (see e.g. \cite{Qian91})
\begin{equation} 
\overline{\Delta Z^2(k,w)} = 
\sum_{i=1}^{N_w-k}\frac{(Z_{i+k} - Z_i) ^{2}}{N_w-k}\,,
\label{eq:MSD-W}
\end{equation}
where $\Delta t$ is the time interval between two successive
measurements, $N_w$ the total number of measurements in $w$ and $Z_{i}
= z(i\Delta t)$ indicates the z-coordinate of an attractive bead of
the TF at time $i\Delta t$. The square deviations (\ref{eq:MSD-W}) are
then averaged over all the $M$ sliding windows of all trajectories
such that $N_w\geq k$:
\begin{equation}
\langle \Delta Z^2(k)\rangle = \frac{1}{M}\sum_{w=1}^M 
\overline{\Delta Z^2(k,w)}\;.
\label{eq:MSD}
\end{equation}
Since our TF is perfectly symmetric, the calculation is independent on
the chosen bead apart from statistical fluctuations whose impact can
be minimized by averaging the results of the two equivalent attractive
beads.  This averaged quantity provides an estimation of the mean
square displacement along $z$ over a time interval $k\Delta t$, which
for a diffusive process behaves as
\begin{equation}
\langle \Delta Z^2(k)\rangle=2D_1 \Delta t \, k\,.
\label{eq:MSD_ave}
\end{equation}
The constant $D_1$ is finally obtained by linear regression.

\subsubsection{Computation of exploration efficiency\label{sec:methods:efficiency}}
The efficiency of DNA exploration by the TF can be estimated from the
fraction of DNA sites not yet visited by the TF during the sliding 
$\mathbf{\nu_{sites}}$.  
The procedure we used
follows Ref.~\cite{Givaty2009} and is described below. At the
beginning of each run the counter of newly probed DNA basis is set to
zero. Then, at the beginning of every sliding window within the run,
each DNA bead is marked with a flag "zero".  Then if the DNA bead
closest to one of the TF attractive bead is within a distance of
1.2$\sigma$ from the latter, the corresponding flag is switched to
``1'' and the counter of probed sites increased by 1.  At the end of
each sliding event, when the TF detaches from DNA, the counter is
normalized to the number of DNA sites to get the fraction of sites
explored in that sliding event. Then the flag vector is reset to zero.
The overall fraction of visited sites is just the sum of the fractions
of sites visited in all the sliding windows in each run. This quantity
is then averaged over all runs to yield $\mathbf{\nu_{sites}}$.  
The strategy to reset the flag vector
to zero is motivated by the assumption that when the TF detaches from
DNA, it is likely to reassociate to a completely uncorrelated sequence
exploring a completely new patch of DNA.

\section{RESULTS \label{Sec:results}}

Experimental studies of facilitated diffusion {\em in vitro} are
conducted by varying the salt concentration which influences the
occurrence of the different transport modes
\cite{Winter81a,Winter81b,elf2007probing}.  As we shall see, in our
model a similar behavior is obtained by changing the parameter
$\sigma$, the equilibrium distance between TF and DNA helix.
An increase in the salt concentration enhances the screening of
electrostatic interactions and thus increases 
the TF-DNA equilibrium distance, which in our model is controlled by 
$\sigma$.

%===========================================================================
\subsection{Statistics of sliding, hopping and jumping
\label{sec:Statistics}}
%===========================================================================

%----------------------------------------------------------------
\begin{figure}[b!]
\centering
\includegraphics[width=0.5\textwidth]{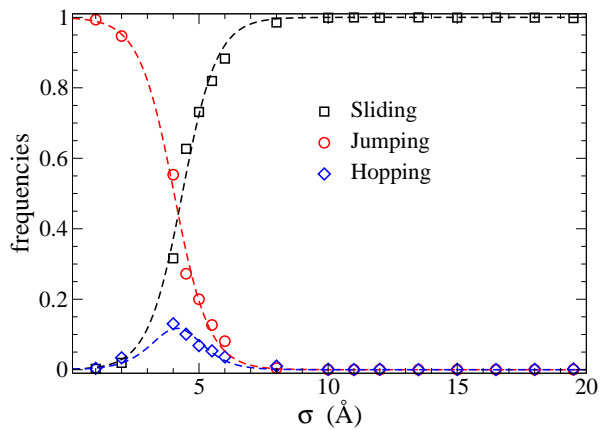}
\caption{Transport modes statistics. Frequencies of sliding (black
  squares), jumping (red circles) and hopping (blue diamonds). The
  dashed lines are just a guide for the eye and have been obtained by
  fitting the data via suitable sigmoid functions. Data are obtained
  by averaging over $50$ runs each lasting $T=10^6$ time 
  units (Methods).\label{fig:1}}
\end{figure}
%----------------------------------------------------------------

\begin{figure}[t!]
\centering
\includegraphics[width=0.5\textwidth]{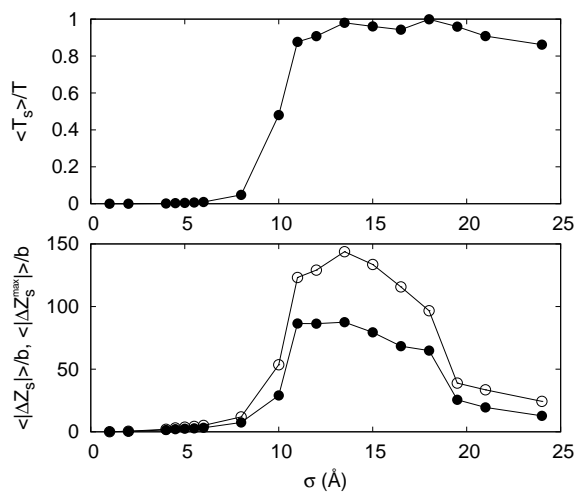}
\caption{Sliding statistics: duration and length. (a) Average duration
  of individual sliding events as a function of $\sigma$ normalized to
  the duration of the single run time window $T=10^6$ code time units.
   Notice that the saturation at
  high $\sigma$ values is a consequence of the fact that sliding
  occurs over the full duration of the simulation run.  (b) Average
  distance $\langle |\Delta Z_s| \rangle$ (full circles) and average
  maximal distance $\langle |\Delta Z^{\mathrm{max}}_s| \rangle$
  (empty circles) covered by the TF in an individual sliding event as
  a function of $\sigma$ and normalized by the base pair distance
  $b$. \label{fig:2}}
\end{figure}

In order to understand the importance of the the different transport
modes (sliding, hopping and jumping) of TFs while interacting with the
DNA, we need first to evaluate their occurrence statistics.  In
Figure~\ref{fig:1}, we show the empirical transport-mode frequencies
of occurrence measured in simulations. The various modes were
identified and analyzed as discussed in \textit{Methods}. We find that
sliding and jumping frequencies follow a sigmoidal profile with a
prevalence of 3D-diffusion at low $\sigma$ and a dominance of sliding
at high $\sigma$.  Hopping events are rare for every value of $\sigma$
except for a small hump around $\sigma \approx 4$\AA. This a likely
consequence of the strongly confining features of the 12-10 LJ
interactions.  When the equilibrium distance between TF and DNA,
$\sigma$, is too small complete dissociation with enduring jumps are
the most probable events. Conversely, for large $\sigma$ values the TF
tends to linger bound to the DNA with sliding events lasting for about
the entire duration of the simulation runs (Fig~\ref{fig:2}a). Hopping
events are statistically significant only at the transition between
the jumping dominated and sliding dominated regimes.
%----------------------------------------------------------------
%----------------------------------------------------------------

Sliding is the most relevant transport mode in determining the TF-DNA
interaction, as only when in sliding the TF can actually probe the
sequence of nucleotides of the DNA.  We thus studied the average
distance along the DNA axis, $\langle |\Delta Z_s| \rangle$, explored
by the TF during a single sliding event. Due to the random walk
character of sliding, in the interval between the times of attachment
and detachment the TF might have moved past the future point of
detachment. Therefore, we also measured the maximal distance from the
point of attachment reached by the TF within the time of detachment,
$\langle |\Delta Z^{\mathrm{max}}_s| \rangle$.  Both $\langle |\Delta
Z_s| \rangle$ and $\langle |\Delta Z^{\mathrm{max}}_s| \rangle$,
normalized by the base pair distance $b$, are shown in
Fig.~\ref{fig:2}b. These quantities provide a proxy of the number of
basis probed between the point of association to the DNA and the
subsequent point of detachment from it. As one can see when sliding is
dominating the number of bases probed by the TF can be as large as
$\sim 150$. However, for values of $\sigma$ larger than $18-20\AA$
while sliding remains the prevailing mechanism of transport
(Fig.~\ref{fig:1} and \ref{fig:2}a), the number of probed bases drops
dramatically. This behavior will be rationalized later while
investigating the behavior of the one-dimensional diffusion constant
$D_{1}$.

\subsection{Geometry of sliding TF trajectories}

%----------------------------------------------------------------
\begin{figure}[b!]
\centering
\includegraphics[width=0.5\textwidth]{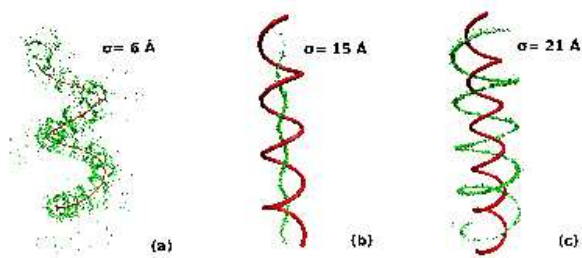}
\caption{Roto-translation of the TF during sliding. The green dots
  represent the positions on one of the attractive TF beads for each
  frame satisfying the sliding condition. The trajectory traced by the
  TF is overlaid to the structure of the DNA chain (red). Panels a-c
  refer to three representative $\sigma$ values as labeled in the
  figure. \label{fig:3}}
\end{figure}
%----------------------------------------------------------------

\begin{figure}[t!]
\centering
\includegraphics[width=0.5\textwidth]{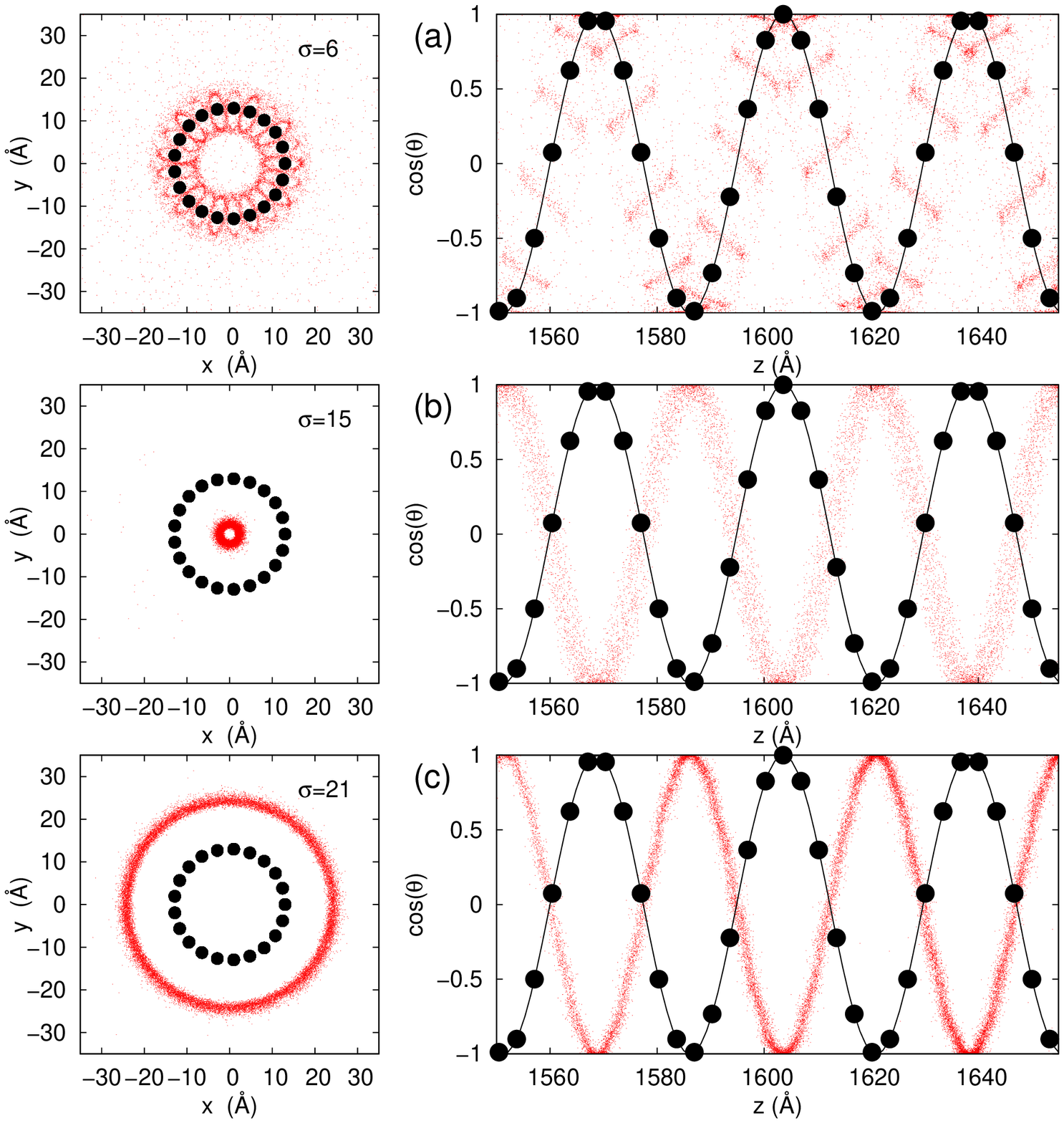}
\caption{Two dimensional projections (red dots) of the TF-bead
  trajectories of Fig.~\ref{fig:3} for $\sigma=6$\AA\; (a), $15$\AA\; (b) and
  $21$\AA\; (c).  Left panels refer to the projection onto the
  $(x,y)$-plane.  Right panels show the the cosine of the angle of
  rotation of the TF around the $z$-axis (DNA axis) as a function of
  the position along the $z$-axis itself.  The black curves correspond
  to the DNA helix $\cos(2\pi z/P)$. Black full circles identify the
  DNA beads. \label{fig:4}}
\end{figure}

We now focus on the geometrical properties of TF trajectories during
sliding. The basic phenomenological features are illustrated in
Figs.~\ref{fig:3} and \ref{fig:4}, showing the positions of one of the
attractive beads (i.e. bead-1 or bead-3) during sliding, for three
representative values of $\sigma$.  Figure~\ref{fig:3} shows a
three-dimensional view, whilst, Figure~\ref{fig:4} displays two
dimensional projections. Denoting with $(x,y,z)$ the Cartesian
coordinates of the bead position, the left panels show the projection
onto the $(x,y)$-plane, transversal to the DNA axis, while the right
ones the cosine of the angle of rotation around the DNA-axis as a
function of the position along the DNA axis,
i.e. $(z,x/\sqrt{x^2+y^2})=(z,\cos(\theta))$.
%----------------------------------------------------------------
%----------------------------------------------------------------

For all values of $\sigma$ we found that, during sliding, the TF
traces the helical path of the DNA as clear from the 3D-plots
(Fig.~\ref{fig:3}), so that diffusion along the DNA chains proceeds
with a characteristic roto-translation as suggested 
by experimental studies \cite{Blainey2009,Liu2008,Kochaniak2009,Lin2009,
Tafvizi2008}, see also the review \cite{Gorman_REV}.
   However, some differences are
observed at varying $\sigma$, as discussed in the following.

For $\sigma=6$\AA\;, the TF traces circular orbits orthogonal to the
helix contour around each DNA bead so that the overall motion draws a
super-helical trajectory (Fig.~\ref{fig:3} and \ref{fig:4}a-left).  In
this case the envelope path is in phase with the DNA helix, as
demonstrated by the behavior of the points representing $\cos(\theta)$
vs $z$ (Fig.~\ref{fig:4}a-right) which accumulate around the curve
$\cos(2\pi z/P)$ ($P$ being the DNA-helix pitch).  For smaller values
of $\sigma$ sliding becomes less frequent but always in phase with the
DNA helix (not reported). For larger values of $\sigma$, the TF
sliding beads tread an helix in antiphase with respect to the DNA
helix (Figs.~\ref{fig:3} and \ref{fig:4}b-c (right)), meaning that the
TF recognition domains reside in the DNA groove.  As far as the
distance from the DNA axis is concerned, for $\sigma=6$\AA\; the TF
bead moves both in and out the DNA helix (Fig.~\ref{fig:4}a-right),
while it remains well inside and outside it for $\sigma=15$\AA\; and
$21$\AA\;, respectively (as shown in (Fig.~\ref{fig:4}b-c (left)).  As
we will show in the next section these observations will be key to
understand the behavior of the one-dimensional diffusion coefficient
$D_1$.

%----------------------------------------------------------------
\begin{figure}[t!]
\centering
\includegraphics[width=0.5\textwidth]{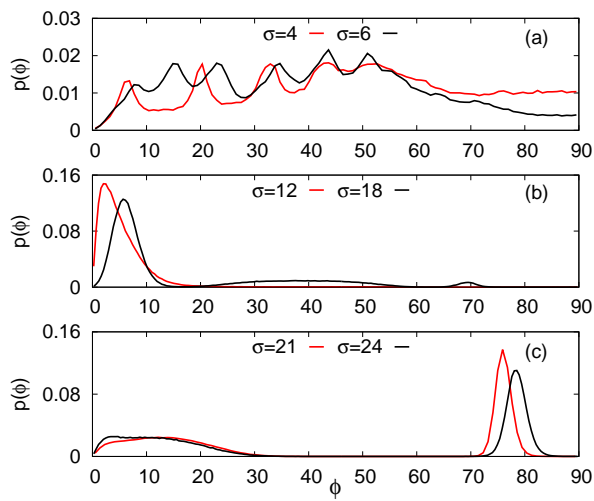}
\caption{Probability density function $p(\phi)$ of the orientation
  angle $\phi$ of the TF with respect to DNA axis. (a), (b) and (c)
  for small, intermediate and high values of $\sigma$ as
  labeled. Notice the difference in the $y$-axis scale between panel
  (a) and (b-c). \label{fig:5}}
\end{figure}
%----------------------------------------------------------------

We complete the description of the geometrical aspects of TF sliding
motion by discussing its orientation with respect to DNA.  Given the
TF triangular geometry, its orientation can be characterized in terms
of the angle $\phi$ between the segment joining bead-1 and bead-3 and
the DNA-axis. Since the dynamics is fully symmetric under the exchange
of bead-1 and bead-3, we can restrict the angle in $[0:90]$
degrees. Measurements are made only when the TF is associated to DNA,
i.e. in the sliding windows.  Figure~\ref{fig:5} shows that the
orientation statistics depends on $\sigma$. For $\sigma<6$\AA\;
(Fig.~\ref{fig:5}a), sliding occurs very rarely and typically only one
of the attractive bead is in contact with DNA, this can be appreciated
from Figs.~\ref{fig:3}a and \ref{fig:4}a-left where the spots out of
the superhelical path corresponds to the instants in which the TF is
in sliding but with one of the beads not attached to DNA. As a
consequence, the probability density, $p(\phi)$, is rather broad with
no preferential orientation. The peaks observed in the figure result
from the loops TF makes around DNA beads.  For intermediate values of
$\sigma$ ($6$\AA$<\sigma<18$\AA\; Fig.~\ref{fig:5}b), the TF describes
an helical path at the interior of the DNA helix (central panel of
Fig.~\ref{fig:3} and Fig.~\ref{fig:4}b) and the $p(\phi)$ takes on a
well defined peak around $\phi=0^o$ meaning that the TF slides along
DNA in parallel orientation, with bead-1 and -3 residing in two
consecutive grooves of the DNA.  As we shall discuss below this
appears to be the fastest TF-DNA configuration in terms of diffusive
properties.  For $\sigma>18$\AA\; (Fig.~\ref{fig:5}c) the peak around
zero broadens and a new peak appears between $70^o$ and $80^o$. These
features signal that now the attractive beads, while performing an
helical motion outside the DNA helix, flip between a parallel
orientation with respect to the DNA axis and an almost orthogonal one,
whereby the two beads straddle the helix.

%=================================================================
\subsection{One-dimensional diffusion coefficient}
%=================================================================

To characterize the TF sliding along the DNA, we estimated from the
runs the one-dimensional diffusion coefficient $D_1$ from the linear
behavior of the mean square displacement along the DNA-axis, see
Eq.~(\ref{eq:MSD_ave}) in \textit{Methods}.

\begin{figure}[t!]
\centering
\includegraphics[width=0.5\textwidth]{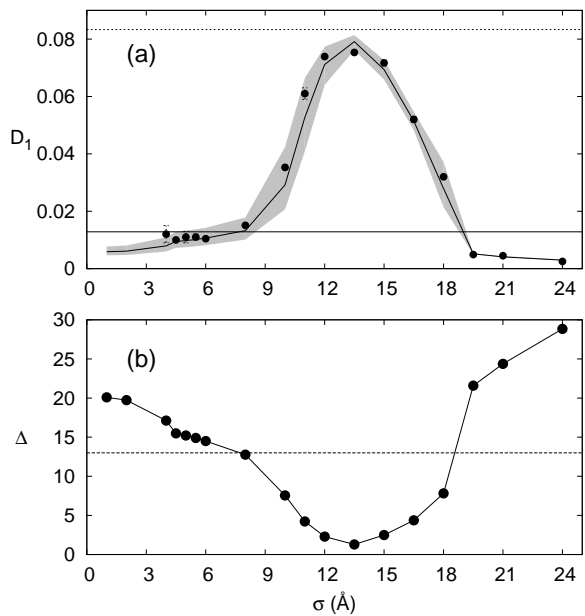}
\caption{Sliding diffusion coefficient. (a) Dependency of the
  diffusion constant along the DNA axis, $D_1$, on $\sigma$. 
  Filled circles denote the measured $D_1$ from simulations; 
  the solid line interpolating data is formula (\ref{eq:D1_theo}) 
  with $\Delta$ being the average TF-DNA distance from numerical 
  simulations; the shaded region is bracketed by the values 
  (\ref{eq:D1_theo}) computed by replacing
   $\Delta$ with $\Delta \pm s_\Delta$, where $s_\Delta$ is the standard
   deviation of the TF-distance from the z-axis.     
  For a
  comparison, the top horizontal dashed-line displays the free diffusion
  coefficient $D_3=k_BT/(3\gamma)$ for the three-bead TF and the bottom 
  solid horizontal line shows formula (\ref{eq:D1_theo}) evaluated at 
  $\Delta=\varrho$.
  (b) Average distance, $\Delta$, of bead-1 and -3 from the
  DNA-axis. The dashed line denote the DNA radius $\varrho$, points
  below this line corresponds to situations in which the beads are on
  average inside the DNA helix.
  \label{fig:6}}
\end{figure}

The results reported in Fig.~\ref{fig:6}a show the dependence of $D_1$
on $\sigma$. The diffusion constant displays small values for both
large and small $\sigma$, while exhibiting a bump for intermediate
values. Figure~\ref{fig:6}b shows the average distance $\Delta$ of the
attractive beads (1 and 3) from the z-axis.  A negative correlation
between $D_1(\sigma)$ and $\Delta(\sigma)$ is apparent indicating that
high values of $D_1$ require a TF deeply embedded into the DNA
groove as highlighted by the horizontal line that marks the DNA radius
$\Delta = \varrho$.

%----------------------------------------------------------------
%----------------------------------------------------------------

The observed behavior of the diffusion constant, $D_1$ can be
rationalized by a simple phenomenological argument based on the
geometrical properties of TF sliding motion that have been
characterized in the previous section.  Basically, during sliding, TF
beads diffuse drawing an helical path (Fig~\ref{fig:3}) at distance
$\Delta$ from the DNA axis (Fig.~\ref{fig:6}b ). Such helical path has
the same pitch $P=hb$ of the DNA helix with, possibly, a phase shift
(Fig~\ref{fig:4}b-c), which is inessential for the following
derivation.  This scenario occurs for $\sigma$ large enough
(Fig.~\ref{fig:3}b-c). For smaller $\sigma$ the path drawn by the TF
is slightly more complex (Fig.~\ref{fig:3}(a)), but on average still
helical.  Assuming an ideal helical motion, the displacement along the
axis, $\delta z$, is linked to the arc-length of a curvilinear
displacement, $\delta \ell$, along the helix by the formula
\begin{equation}
{\delta z}= \frac{\delta \ell}{\sqrt{1+(2\pi \Delta/P)^2}}\,.
\label{eq:conversion}
\end{equation}
From Einstein equation for a three-atom molecule, like our TF, the
mean square curvilinear displacement along the helical path is
$\langle \delta \ell(t)^2\rangle=2D_3 t$ with $D_3=k_BT/(3\gamma)$,
while along the $z$-axis we have $\langle \delta z(t)^2\rangle=2D_1
t$. Then, using Eq.~(\ref{eq:conversion}) to convert displacements
along the helix to displacements along the DNA axis yields
\begin{equation}
\label{eq:D1_theo}
D_1 = \frac{D_3}{1+ (2\pi \Delta/P)^2}=
\frac{k_B T}{3\gamma\left[1+ (2\pi \Delta/P)^2\right]}\,,
\end{equation}
which relates the diffusion constant along the z-axis to the
geometrical properties of the helical path followed by the TF.  The
shaded region in Fig.~\ref{fig:6}a is bracketed by the upper and lower
bounds of $D_1$ obtained using Eq.~(\ref{eq:D1_theo}) by replacing
$\Delta$ with $\Delta \pm s_\Delta$, where $s_\Delta$ is the standard
deviation of the TF-distance from the z-axis.  The region accurately
brackets the simulation data supporting the reliability of the
prediction (\ref{eq:D1_theo}). For instance, if the TF were diffusing
on an helix with radius equal to that of the DNA it would correspond
to a $D_1$ with the value marked by the solid line in
Fig.~\ref{fig:6}a.

It is interesting to observe that Eq.~(\ref{eq:D1_theo}) is consistent
with the theoretical prediction of Bagchi et al.~\cite{Bagchi2008}
based on the computation of the translational friction induced by the
TF helical track along the DNA. To obtain the connection one should
neglect the friction contribution of TF self-rotation predicted by
Schurr~\cite{Schurr1979} which is not relevant to our model.

The ability of Eq.~(\ref{eq:D1_theo}) to quantitatively explain the
behavior of simulated sliding diffusion constant suggests that,
within our model, $D_1$ is mainly determined by the geometrical
properties of sliding path.  In other terms, the DNA-geometry
conspires with the interaction potential to constrain the TF to
diffuse along an helical path without being much influenced by
possible potential barriers, indeed the derivation was based on the
free diffusion coefficient $D_3$.  Essentially the effect of the
interaction potential is embodied in the fact that $\Delta$ in
Fig.~\ref{fig:6}b results to be a non trivial function of $\sigma$. In
principle, the interaction potential between TF and DNA depends on the
nucleotide sequence, so that diffusion is modified by the presence of
a rugged energy landscape \cite{Mirny2009,Sheinman2012}. This effect
typically depresses the diffusion, in particular it will affect the
value of $D_3$ used in the argument above by decreasing it by a factor
$\propto \exp[-(E_v/k_BT)^2]$, $E_v$ being the standard deviation 
of the TF-DNA
(now disordered) binding energy.  However, experimental data suggest
that this effect, when present, is very small with $E_v\leq
1k_BT-2k_BT$ \cite{Mirny2009}. Of course, the model we introduced can
be easily generalized to include sequence heterogeneity.

%===================================================================
\subsection{Search efficiency
\label{sec:Efficiency}}
%===================================================================

Even in a scenario of non-specific TF-DNA interaction it is
interesting to quantify the efficiency of sequence exploration during
sliding.  Following Ref.~\cite{Givaty2009}, we estimate the
exploration efficiency in terms of ``probed positions'', i.e.  we
measure the fraction of new sites, $\nu_{\mathrm{sites}}$, visited by
the sliding TF, see \textit{Methods}.

%----------------------------------------------------------------
\begin{figure}[t!]
\centering
\includegraphics[width=0.5\textwidth]{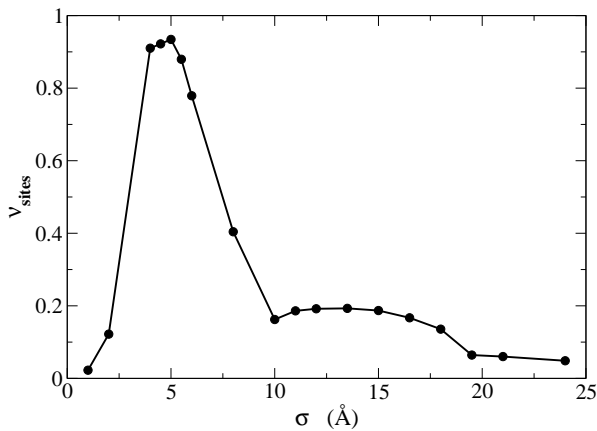}
\caption{Fraction $\nu_{sites}$ of positions probed by the TF during
  sliding as a function of $\sigma$.}
\label{fig:7}. 
\end{figure}
%----------------------------------------------------------------

In Figure~\ref{fig:7} we show the fraction $\nu_{\mathrm{sites}}$ as a
function of $\sigma$. The exploration efficiency displays a well
pronounced peak in the range $4$\AA $< \sigma <$ $5.5$\AA. At a first
sight, this result may look surprising as for such values of $\sigma$
the diffusion constant $D_1$ is rather smaller than its maximum value
attained at $\sigma \approx 13$\AA\, (Fig.~\ref{fig:6}). However, this
behavior represents the essence of facilitated diffusion whereby slow
sliding can be compensated by frequent jumping and hopping. Indeed, a
direct comparison between Fig~\ref{fig:7} and \ref{fig:1} reveals that
the search efficiency peaks in the region where hopping is maximal and
jumping/sliding events have comparable frequencies. In other terms,
the possibility to realize an efficient search through the DNA chain
to localize as quick as possible the DNA target sequence requires a
suitable interplay of all transport modes. The result shown in
Fig~\ref{fig:7} is in qualitative agreement with those observed in
Ref.~\cite{Givaty2009}.

%%%%%%%%%%%%%%%%%%%%%%%%%%%%%%%%%%%%%%%%%%%%%%%%%%%%%%%%%%%%%%%%%%%
\section{DISCUSSION and CONCLUSIONS}
%%%%%%%%%%%%%%%%%%%%%%%%%%%%%%%%%%%%%%%%%%%%%%%%%%%%%%%%%%%%%%%%%%%
In this work we performed Molecular Dynamics simulations of
facilitated diffusion using a very simplified model. We represented
DNA as a single helical chain of beads frozen in the standard
conformation of B-DNA.  To capture the main features of typical
prokaryotic homodimeric Transcription Factors (TFs) that target
palindromic DNA sequences~\cite{BrandenTooze}, the TF was represented
as three-bead triangular structure, where the first and last bead
correspond to the binding regions whereas the central bead to the
scaffold.  Our model does not include electrostatics and the TF-DNA
interactions are modeled through a Lennard-Jones potential of well
depth $\epsilon$ and equilibrium distance $\sigma$.  In our
simulations, $\epsilon$ is kept constant while exploring a wide range
of $\sigma$ values.

Our simulations show that the DNA-TF equilibrium distance $\sigma$
crucially affects the dynamics of the TF. For small $\sigma$ the TF
spends most of its time in 3D-diffusion. At intermediate values of
$\sigma$ a sharp transition occurs with a drop in the jumping
frequency and an abrupt increase in the sliding frequency paralleled
by the appearance of a hump in the hopping frequency and an increase
of the $D_{1}$ diffusion constant. The shape of the trajectory traced
by the TF is also very sensitive to $\sigma$. For intermediate
$\sigma$ the TF forms circular orbits orthogonal to the DNA contour
creating a super-helical path.  For larger $\sigma$ the TF traces an
helical trajectory in phase with the DNA groove. Both the attractive
beads of the TF are accommodated at the bottom of the groove imparting
a parallel orientation to the DNA axis. Finally, for even larger
$\sigma$ the coils of this helical path in antiphase with the DNA
helix become wide enough to cause a drop in the $D_{1}$ diffusion
constant. The behavior of the $D_{1}$ constant was explained by a
simple geometric argument based on the projection of the mean square
displacement of the TF trajectory onto the DNA axis.  The expression
that we derived is similar to the one introduced in
Ref.~\cite{Bagchi2008} save for the self-rotation frictional
contribution \cite{Schurr1979}, that in our simulations is not
relevant. We did not find any apparent dependence of $D_1$ on possible
energetic barriers. However, the latter enter the expression of the
diffusion constant in an implicit way by setting the average distance
of the TF from DNA.

An interesting feature of our model is that, despite the crude
approximations, it could reproduce a number of known phenomenological
patterns. For instance, for intermediate values of $\sigma$, the TF
always remains deeply buried into the groove of the DNA molecule with
a parallel orientation with respect to DNA axis.  This result is in
agreement with a Circular Dichroism study by Johnson \emph{et
  al}~\cite{Johnson94} showing that the TF interaction with unspecific
DNA sequences is sufficient to induce the structuring of typical
DNA-binding motifs, and is confirmed by recent NMR analyzes by Iwahara
\emph{et al}~\cite{Iwahara2006} revealing that protein HoxD9 interacts
with non-specific binding sites using the same interface employed for
the recognition of the specific target site. This result, also
consistent with Givaty and Levy findings~\cite{Givaty2009}, might have
far reaching biological implications~\cite{Marcovitz2011} suggesting
the existence of only a very low barrier separating the Search and
Recognition states postulated in Ref.~\cite{Slutsky2004}.

It is also interesting to notice that in our model the translational
move of the TF during sliding is always coupled to rotation around the
DNA axis induced by the helical path, which is either in phase with
the DNA strand or with its groove. This result is consistent with
single molecule fluorescence tracking assays performed by Blainey
\emph{et al}~\cite{Blainey2009} for the calculation of $D_{1}$ of
labeled human oxoguanine DNA glycosylase (hOgg1). This study was based
on the observation that in the case of pure translation the $D_{1}$
coefficient depends on the TF radius $R$ like $1/R$ while in the case
of roto-translation $D_{1} \propto 1/[(4/3)R^{3} + R(R_{OC})^{2}]$
(with $R_{OC}$ being the distance of the TF from the DNA axis,
i.e. $\Delta$ in our notation, see Fig.~\ref{fig:6}b) so that purely
translational and roto-translational sliding can be
discriminated. This work, along with other recent
studies~\cite{Liu2008, Kochaniak2009,Lin2009} extending the analysis
of sliding to several other proteins, supports the suggestive idea that the
coupling between rotation and translation might be a feature 
shared, at least, by a group of TFs.

As an overall conclusion, our model, despite its crude approximations,
turns out to reproduce fairly well a number of experimental
patterns. This represents an \emph{a posteriori} validation of the two
key elements of our scheme, namely the helical topology of the DNA
molecule and a TF-DNA interaction potential with a well localized
minimum and a short tail.  This extensive model validation suggests
its viability in investigating more complex aspects of facilitated
diffusion such as the influence of molecular crowding and DNA
flexibility.

\ack
We thank A. Vulpiani for discussions. FC and MC acknowledge support
from MIUR PRIN-2009PYYZM5.

\section*{References} 
%\bibliographystyle{unsrt}
%\bibliography{biblio_DNA_and_TF}

\begin{thebibliography}{90}

\bibitem{Alberts}
B.~Alberts, A.~Johnson, J.~Lewis, M.~Raff, K.~Roberts, and P.~Walter.
\newblock {\em Molecular biology of the cell}.
\newblock Garland science, Taylor \& Francis Group, New York, USA, 5th edition,
  2007.

\bibitem{Riggs70}
A.D. Riggs, S.~Bourgeois, and M.~Cohn.
\newblock The lac repressor-operator interaction: Iii. kinetic studies.
\newblock {\em J. Mol. Biol.}, 53:401--417, 1970.

\bibitem{Berg81}
O.G. Berg, R.B. Winter, and von Hippel~P.H.
\newblock Diffusion-driven mechanisms of protein translocation on nucleic
  acids. 1. models and theory.
\newblock {\em Biochemistry}, 20:6929--6948, 1981.

\bibitem{halford2004}
S.~E. Halford and J.~F. Marko.
\newblock How do site-specific dna-binding proteins find their targets?
\newblock {\em Nucleic Acids Res.}, 32(10):3040--3052, 2004.

\bibitem{Slutsky2004}
M.~Slutsky and L.A. Mirny.
\newblock Kinetics of {protein-DNA} interaction: facilitated target location in
  sequence-dependent potential.
\newblock {\em Biophys. J.}, 87:4021--4035, 2004.

\bibitem{Sheinman2012}
M.~Sheinman, O.~B\'{e}nichou, Y.~Kafri, and R.~Voitouriez.
\newblock Classes of fast and specific search mechanisms for proteins on {DNA}.
\newblock {\em Rep. Prog. Phys.}, 75:026601 (33pp), 2012.

\bibitem{Chu2009}
D.~Chu, N.R. Zabet, and J.~Mitavskiy.
\newblock Models of transcription factor binding: Sensitivity of activation
  functions to model assumptions.
\newblock {\em J. Theor. Biol.}, 257:419--429, 2009.

\bibitem{Zabet2012}
N.R. Zabet and B.~Adryan.
\newblock A comprehensive computational model of facilitated diffusion in
  prokaryotes.
\newblock {\em Bioinformatics}, 28:1517--1524, 2012.

\bibitem{Elf2007}
J.~Elf, G.-W. Li, and X.S. Xie.
\newblock Probing transcription factor dynamics at the single-molecule level in
  a living cell.
\newblock {\em Science}, 316:1191--1194, 2007.

\bibitem{Wang2006}
Y.M. Wang, R.H. Austin, and E.C. Cox.
\newblock Single molecule measurements of repressor protein 1d diffusion on
  {DNA}.
\newblock {\em Phys. Rev. Lett.}, 97:048302, 2006.

\bibitem{Graneli2006}
A.~Graneli, C.C. Yeykal, R.B. Robertson, and E.C. Greene.
\newblock Long-distance lateral diffusion of human rad51 on double-stranded
  {DNA}.
\newblock {\em Proc. Natl. Acad. Sci. USA}, 103:1221--1226, 2006.

\bibitem{Benichou2008}
O.~B\'{e}nichou, C.~Loverdo, and R.~Voitouriez.
\newblock How gene colocalization can be optimized by tuning the diffusion
  constant of transcription factors.
\newblock {\em Europhys. Lett.}, 84:38003, 2008.

\bibitem{Laughton2011}
C.~Laughton and S.A. Harris.
\newblock The atomistic simulation of {DNA}.
\newblock {\em Comput. Mol. Sci.}, 1:590--600, 2011.

\bibitem{Mirny2009}
L.~Mirny, M.~Slutsky, Z.~Wunderlich, A.~Tafvizi, J.~Leith, and A.~Kosmrlj.
\newblock How a protein searches for its site on dna: the mechanism of
  facilitated diffusion.
\newblock {\em J. Phys. A: Math. and Theor.}, 42(43):434013, 2009.

\bibitem{Brackley2012}
C.A. Brackley, M.E. Cates, and D.~Marenduzzo.
\newblock Facilitated diffusion on mobile {DNA}: configurational traps and
  sequence heterogeneity.
\newblock {\em Phys. Rev. Lett.}, 109:168103, 2012.

\bibitem{Givaty2009}
O.~Givaty and Y.~Levy.
\newblock Protein sliding along {DNA}: dynamics and structural
  characterization.
\newblock {\em J. Mol. Biol.}, 385:1087--1097, 2009.

\bibitem{Bagchi2008}
B.~Bagchi, P.~C Blainey, and X~S. Xie.
\newblock Diffusion constant of a nonspecifically bound protein undergoing
  curvilinear motion along dna.
\newblock {\em J. Phys. Chem. B}, 112(19):6282--6284, 2008.

\bibitem{Schurr1979}
J~Michael Schurr.
\newblock The one-dimensional diffusion coefficient of proteins absorbed on
  {DNA}: Hydrodynamic considerations.
\newblock {\em Biophys. Chem.}, 9(4):413--414, 1979.

\bibitem{Gorman_REV}
Jason Gorman and Eric~C Greene.
\newblock Visualizing one-dimensional diffusion of proteins along {DNA}.
\newblock {\em Nature Struct. \& Mol. Biol.}, 15(8):768--774, 2008.

\bibitem{Blainey2009}
P.C. Blainey, L.~Guobin, S.C. Kou, W.F. Mangel, G.L. Verdine, B.~Bagchi, and
  X.S. Xie.
\newblock Nonspecifically bound proteins spin while diffusing along {DNA}.
\newblock {\em Nature Struc \& Mol. Biol.}, 16:1224--1230, 2009.

\bibitem{BrandenTooze}
C.~Branden and J.~Tooze.
\newblock {\em Introduction to protein structure}.
\newblock Garland Publishing, New York, USA, 2nd edition, 1999.

\bibitem{Hinchliffe2008}
A.~Hinchliffe.
\newblock {\em Molecular modelling for beginners}.
\newblock Wiley, Chichester, U.K., 2nd edition, 2008.

\bibitem{Cacciuto2006}
A.~Cacciuto and E.~Luijten.
\newblock Self-avoiding flexible polymers under spherical confinement.
\newblock {\em Nano Lett.}, 6:901--905, 2006.

\bibitem{Berg1977}
O.~G Berg and C.~Blomberg.
\newblock Association kinetics with coupled diffusion. an extension to
  coiled-chain macromolecules applied to the lac repressor-operator system.
\newblock {\em Biophys. Chem.}, 7(1):33--39, 1977.

\bibitem{Melchionna2007}
S.~Melchionna.
\newblock Design of quasi-symplectic propagators for langevin dynamics.
\newblock {\em J. Chem. Phys.}, 127:044108, 2007.

\bibitem{Padding}
J.~T. Padding and A.~A. Louis.
\newblock Hydrodynamic interactions and brownian forces in colloidal
  suspensions: Coarse-graining over time and length scales.
\newblock {\em Phys. Rev. E}, 74:031402, Sep 2006.

\bibitem{Gowers2005}
D.~M Gowers, G.~G Wilson, and S.~E. Halford.
\newblock Measurement of the contributions of 1d and 3d pathways to the
  translocation of a protein along {DNA}.
\newblock {\em Proc. Natl. Acad. Sci. USA}, 102(44):15883--15888, 2005.

\bibitem{Qian91}
H.~Qian, M.P. Sheetz, and E.L. Elson.
\newblock Single particle tracking. analysis of diffusion and flow in
  two-dimensional systems.
\newblock {\em Biophys. J.}, 60:910--921, 1991.

\bibitem{Winter81a}
R~B Winter, O~G Berg, and P~H Von~Hippel.
\newblock Diffusion-driven mechanisms of protein translocation on nucleic
  acids. 3. the escherichia coli lac repressor-operator interaction: kinetic
  measurements and conclusions.
\newblock {\em Biochemistry}, 20(24):6961--6977, 1981.

\bibitem{Winter81b}
R.~B Winter, O.~G Berg, and P.~H Von~Hippel.
\newblock Diffusion-driven mechanisms of protein translocation on nucleic
  acids. 3. the escherichia coli lac repressor-operator interaction: kinetic
  measurements and conclusions.
\newblock {\em Biochemistry}, 20(24):6961--6977, 1981.

\bibitem{elf2007probing}
J.~Elf, G.-W. Li, and X~S. Xie.
\newblock Probing transcription factor dynamics at the single-molecule level in
  a living cell.
\newblock {\em Science}, 316(5828):1191--1194, 2007.

\bibitem{Liu2008}
S.~Liu, E.A. Abbondanzieri, J.W. Rausch, S.F.J. Grice, and X.~Zhuang.
\newblock Slide into action: dynamic shuttling of {HIV} reverse transcriptase
  on nucleic acid substrates.
\newblock {\em Science}, 322:1092--1097, 2008.

\bibitem{Kochaniak2009}
A.B. Kochaniak, S.~Habuchi, J.J. Loparo, D.J. Chang, K.A. Cimprich, J.C.
  Walter, and A.M. van Oijen.
\newblock Proliferating cell nuclear antigen uses two distinct modes to move
  along {DNA}.
\newblock {\em J. Biol. Chem.}, 284:17700--17710, 2009.

\bibitem{Lin2009}
Y.~Lin, T.~Zhao, X.~Jian, Z.~Farooqui, X.~Qu, C.~He, A.R. Dinner, and N.F.
  Scherer.
\newblock Using the bias from flow to elucidate single {DNA} repair protein
  sliding and interactions with {DNA}.
\newblock {\em Biophys. J.}, 96:1911--1917, 2009.

\bibitem{Tafvizi2008}
A.~Tafvizi, F.~Huang, J.S. Leith, A.R. Fersht, L.A. Mirny, and A.M. van Oijen.
\newblock Tumor suppressor p53 slides on {DNA} with low friction and high
  stability.
\newblock {\em Biophys. J.}, 95:L01--L03, 2008.

\bibitem{Johnson94}
N.P. Johnson, J.~Lindstrom, W.A. Baase, and P.H. vpn Hippel.
\newblock Double-stranded {DNA} templates can induce alpha-helical conformation
  in peptides containing lysine and alanine. functional implications for
  leucine-zipper and helix-loop-helix transcription factors.
\newblock {\em Proc. Natl. Acad. Sci. USA}, 91:4840--4844, 1994.

\bibitem{Iwahara2006}
J.~Iwahara and G.M. Clore.
\newblock Detecting transient intermediates in macromolecular binding by
  para-magnetic {NMR}.
\newblock {\em Nature}, 440:1227--1230, 2006.

\bibitem{Marcovitz2011}
A.~Marcovitz and Y.~Levy.
\newblock Frustration in {protein-DNA} binding influences conformational
  switching and target search kinetics.
\newblock {\em Proc. Natl. Acad. Sci. USA}, 108:17957--17962, 2011.

\end{thebibliography}

%%%%%%%%%%%%%%%%%%%%%%%%%%%%%%%%%%%%%%%%%%%%%%%%%%%%%%%%%%%%%%%%%%%%%%%%
\appendix \section{}
%%%%%%%%%%%%%%%%%%%%%%%%%%%%%%%%%%%%%%%%%%%%%%%%%%%%%%%%%%%%%%%%%%%%%%%%
Since detailed models of TF-DNA interaction normally include electrostatics
through a Debye-H\"{u}ckel potential (DH), in this Appendix we will show that 
the Lennard-Jones potential (LJ) is flexible enough to account for 
relevant aspects of screened electrostatic interactions.
As customary, in order to include the excluded volume effect and automatically 
remove possible singularities, the original DH potential  
is complemented with a short-range repulsive (see eg. \cite{Givaty2009})
\begin{equation}
V_{el}(r) = 5\epsilon \bigg [\bigg(\frac{\sigma}{r}\bigg)^{12}  
+B \frac{e^{-\lambda r}}{r} \bigg ]\,,
\label{eq:Vel}
\end{equation}
where $\lambda$ is the inverse of the screening length.   
To achieve a mapping with LJ potential, $\lambda$ and $B$ are parameters to be 
adjusted such 
that $V_{el}$ presents the same position $x=\sigma$ and depth 
$V_{min} = -\epsilon$ of LJ minimum. 
It is convenient to rescale the distance $x = r/\sigma$ so that the minimum 
of $V_{LJ}$ potential lies at $x=1$. The parameters $\lambda$ and $B$ are 
obtained by solving the system 
\begin{eqnarray}
\frac{\partial V_{el}}{\partial x} \bigg|_{x=1} = 0 \\ 
 V_{el}(x=1) = -\epsilon  
\end{eqnarray}
Simple algebraic manipulations yield $\lambda \sigma = 9$ 
and $B=6\sigma \exp(\lambda \sigma)/5$ so that the 
interaction potential reads
\begin{equation}
V_{el}(r) = 5\epsilon \bigg [ \bigg ( \frac{\sigma}{r} \bigg )^{12}
- \frac{6}{5} \frac{e^{-9(r/\sigma - 1)}}{r/\sigma} \bigg ]
\end{equation}       
Since $\lambda$ is known to be proportional to the inverse of the 
square root of the salt concentration $C_{s}$, it follows that 
$\sigma \propto \sqrt{C_{s}}$. 

This simple argument shows that given a $V_{el}$ potential, it is always 
possible to determine an approximating (equivalent) LJ potential
characterized by a minimum with the same position and depth. A useful
by-product of the employment of LJ consists in the possibility to readily
locate the putative equilibrium position.  

\end{document}